\documentstyle[multicol,aps,epsf,graphicx]{revtex}
\begin{document}
\title{Nontrivial Velocity Distributions in Inelastic Gases}
\author{P.~L.~Krapivsky$^1$ and E.~Ben-Naim$^2$}
\address{$^1$Center for Polymer Studies and Department of Physics,
Boston University, Boston, MA 02215}
\address{$^2$Theoretical Division and Center for Nonlinear Studies,
Los Alamos National Laboratory, Los Alamos, NM 87545}
\maketitle
\begin{abstract}
We study freely evolving and forced inelastic gases using the
Boltzmann equation. We consider uniform collision rates and obtain
analytical results valid for arbitrary spatial dimension $d$ and
arbitrary dissipation coefficient $\epsilon$. In the freely evolving
case, we find that the velocity distribution decays algebraically,
$P(v,t)\sim v^{-\sigma}$ for sufficiently large velocities. We derive
the exponent $\sigma(d,\epsilon)$, which exhibits nontrivial
dependence on both $d$ and $\epsilon$, exactly. In the forced case,
the velocity distribution approaches a steady-state with a Gaussian
large velocity tail.

\vspace{5pt}
\end{abstract}

\begin{multicols}{2}

The dynamics of granular gases present novel challenges, previously
not encountered in fluid dynamics \cite{pl}.  Specifically, the strong
underlying energy dissipation leads to clustering instabilities and 
strong velocity correlations \cite{gz,soto,blk,bcdr,rbss,slk}. A
series of recent experimental and theoretical studies reveals a rich
phenomenology. In particular, velocities are characterized by
anomalous statistics, sensitive to the details of the driving
conditions, the density, and the degree of dissipation
\cite{ep,ve,lcdkg,ou,nbc,rm,bbrtv,ao}.

Kinetic theory provides a systematic framework for deriving
macroscopic properties from the microscopic collision dynamics
\cite{jr,sg,gs}. Yet, analysis of the corresponding Boltzmann
equation often involves uncontrolled approximations or use of
nearly Maxwellian distributions. Motivated by the latter issue, we
examine both unforced and forced inelastic gases using a
simplified Boltzmann equation. Specifically, we shall employ
Maxwell's collision rate which is proportional to the typical
velocity rather than the relative velocity\cite{e}, and show that
in general, it is analytically tractable.

Previous studies reveal that this kinetic theory leads to
interesting behaviors in the freely evolving case. In one
dimension, while moments of the velocity distribution exhibit
multiscaling \cite{bk}, the velocity distribution itself still
approaches a scaling form with an algebraic large velocity tail
\cite{bmp}. An algebraic tail was also found numerically in two
dimensions \cite{bmp}.  Here, we show analytically that in
arbitrary spatial dimension the velocity distribution admits a
scaling solution with an algebraic large velocity tail.  The
corresponding exponent, a root of a transcendental equation,
depends on the spatial dimension and the restitution coefficient.
In the driven case, we find that although the velocity
distribution is non-Maxwellian, it does exhibit a Gaussian tail.

Our starting point is a homogeneous system of identical inelastic spherical
particles in arbitrary spatial dimension $d$.  The mass and the cross-section
are set to unity without loss of generality.  When two particles collide, the
normal component of the relative velocity is reduced by the restitution
coefficient $r=1-2\epsilon$, while the tangential component remains the same.
Denoting by ${\bf n}$ the unit vector connecting the centers of the colliding
particles, the post-collision velocities ${\bf v}_{1,2}$ are given by a
linear combination of the precollision velocities ${\bf u}_{1,2}$,
\begin{equation}
\label{inel}
{\bf v}_{1,2}={\bf u}_{1,2}\mp(1-\epsilon)\,({\bf g}\cdot {\bf n})\,{\bf n},
\end{equation}
with the relative velocity ${\bf g}={\bf u}_1-{\bf u}_2$.  The energy
dissipated in a collision equals $\Delta E=-\epsilon(1-\epsilon)({\bf
g}\cdot{\bf n})^2$.  When $\epsilon=0$ collisions are elastic, while
when $\epsilon=1/2$ collisions are perfectly inelastic with maximal
energy dissipation.

Let $P({\bf v},t)$ be the normalized density of particles with velocity ${\bf
  v}$ at time $t$. Assuming molecular chaos or perfect mixing, we arrive at
the following Boltzmann equation for the velocity distribution
\end{multicols}
\begin{equation}
\label{be}
{\partial P({\bf v},t)\over \partial t} = \langle g\rangle
\int d{\bf n}\int d{\bf u}_1 \int d{\bf u}_2\,\,P({\bf u}_1,t)\,
P({\bf u}_2,t)\,\Big\{\delta\big[{\bf v}-{\bf u}_1 +(1-\epsilon)
({\bf g}\cdot {\bf n})\,{\bf n}\big] -\delta({\bf v}-{\bf u}_1)\Big\}.
\end{equation}
\begin{multicols}{2}
\noindent The collision rate in the Maxwell approximation represent
the typical velocity scale, $\langle g\rangle=\sqrt{T}$, with
$T={1\over d}\langle v^2\rangle$ the ``granular temperature'' or the
average kinetic energy per degree of freedom (for hard spheres, the
collision rate equals the actual relative velocity ${\bf g}\cdot {\bf
n}$). This evolution equation naturally describes a stochastic process
where randomly chosen pairs of particles undergo inelastic collisions
according to (\ref{inel}) with a randomly chosen impact direction $\bf
{n}$. In writing Eq.~(\ref{be}) we tacitly ignored the restriction
${\bf g}\cdot {\bf n}>0$ on the integration range, because the
integrand obeys the reflection symmetry ${\bf n}\to -{\bf n}$.  The
integration measure should be normalized, $\int d{\bf n}=1$.

In the absence of energy input the system ``cools'' indefinitely according to
Haff's law \cite{pkh}. Indeed, the time dependence of the temperature is
found from the Boltzmann equation (\ref{be}) to give ${d\over dt}T=-\lambda
T^{3/2}$ with \hbox{$\lambda=2\epsilon(1-\epsilon)\int d{\bf n}\, n_1^2$}
(the first axis was conveniently chosen to be parallel to ${\bf g}$).  Since
$n_1^2+\ldots+n_d^2=1$ and $\int d{\bf n}=1$ one has
$\lambda=2\epsilon(1-\epsilon)/d$. Thus, the temperature decays according to
$T(t)=T_0\,(1+t/t_0)^{-2}$, with the initial temperature $T_0$ and the
characteristic time scale $t_0=d/\big[\epsilon(1-\epsilon)\sqrt{T_0}\,\big]$.
The temperature quantifies velocity fluctuations. In the following, we focus
on the natural case of isotropic velocity distributions. We will show that
asymptotically, the temperature represents the only relevant velocity scale
as the velocity distribution approaches the scaling form
\begin{equation}
\label{pscl}
P({\bf v},t)\sim {1\over T^{d/2}}{\cal
P}\left({v\over \sqrt{T}}\right).
\end{equation}

Given the convolution structure of the Boltzmann equation (\ref{be}),
we introduce $F({\bf k}, t)$, the Fourier transform of the velocity
distribution function, \hbox{$F({\bf k}, t)=\int d{\bf v}\,e^{i{\bf k}\cdot
{\bf v}}\,P({\bf v},t)$}.  Applying the Fourier transform to
Eq.~(\ref{be}) and integrating over the velocities gives
\begin{equation}
\label{fkt}
{1\over \sqrt{T}}{\partial \over \partial t}F({\bf k},t)+F({\bf k},t)=
\int d{\bf n}\,F\left[{\bf k}-{\bf q},t\right]F\left[{\bf q},t\right],
\end{equation}
with ${\bf q}=(1-\epsilon)({\bf k}\cdot {\bf n})\,{\bf n}$.  This rate
equation reflects the momentum transferred between particles during
collisions.  We seek an isotropic scaling solution for the Fourier
transform, the equivalent of (\ref{pscl}),
\begin{equation}
\label{fscl}
F({\bf k},t)=\Phi\left(k^2 T\right),
\end{equation}
with $k\equiv |{\bf k}|$.  In the $k\to 0$ limit, the Fourier
transform, $F({\bf k},t)\cong 1-{1\over 2}\,k^2\, T$, implies that
the first two terms in the Taylor expansion of the corresponding
scaling function $\Phi(x)$ are universal, $\Phi(x)\cong 1-{1\over
2}x$.

Let us first consider the simpler 1D case.  Equation (\ref{fkt}) reduces to
\hbox{${1\over \sqrt{T}}{\partial \over \partial t}F(k,t)+F(k,t)= F[\epsilon k,
  t]\,F[k-\epsilon k, t]$} and the scaling function (\ref{fscl})
satisfies\cite{bk}
\begin{equation}
\label{fx1}
-\lambda \,x\,\Phi'(x)+\Phi(x)
=\Phi\left[\epsilon^2 x\right]\,\Phi\left[(1-\epsilon)^2 x\right].
\end{equation}
This equation admits a very simple solution \cite{bmp}
\begin{equation}
\label{fsol}
\Phi(x)=\left[1+\sqrt{x}\,\right]\,e^{-\sqrt{x}}.
\end{equation}
The inverse Fourier transform gives the scaling function of the velocity
distribution as a squared Lorentzian
\begin{equation}
\label{Pvt} {\cal P}(w)={2\over \pi}{1\over (1+w^2)^2}.
\end{equation}
Therefore, the form of the scaling solution (\ref{Pvt}) is universal
as it is independent of the dissipation degree.  Another
important feature is the algebraic tail of the velocity distribution,
${\cal P}(w)\sim w^{-4}$ as $w\to\infty$.

We now return to the general $d$-dimensional case.  Substituting
(\ref{fscl}) into (\ref{fkt}) and using the temperature cooling
equation ${d\over
  dt}T=-\lambda T^{3/2}$ one finds that the scaling function $\Phi(x)$
satisfies $\Phi(x)-\lambda x\Phi'(x)=\int\! d{\bf n}\,\Phi(\xi
x)\Phi(\eta x)$, where $\xi=1-(1-\epsilon^2)n_1^2$ and
$\eta=(1-\epsilon)^2n_1^2$.  To integrate over the impact
direction ${\bf n}$ we use spherical coordinates and treat the
first axis as the polar axis, $n_1=\cos \theta$.  The
$\theta$-dependent factor of the measure is $d{\bf n}=N^{-1}(\sin
\theta)^{d-2} d\theta$ with the factor \hbox{$N=\int_0^{\pi} (\sin
\theta)^{d-2} d\theta =B\left({1\over 2}, {d-1\over 2}\right)$}
ensuring proper normalization ($B(a,b)$ is the beta function).
Using $\mu=\cos^2 \theta$ as the integration variable, the above
governing equation for the scaling function reads
\begin{equation}
\label{fx}
-\lambda \,x\,\Phi'(x)+\Phi(x)
=\int_0^1 {\cal D}\mu\,\,\Phi(\xi x)\,\,\Phi(\eta x),
\end{equation}
where $\xi(\mu)=1-(1-\epsilon^2)\mu$ and $\eta(\mu)=(1-\epsilon)^2\mu$.
Additionally, the integration measure was re-written using ${\cal D}\mu$,
defined via $B\left({1\over 2},{d-1\over 2}\right){\cal D}\mu= \mu^{-{1\over
    2}}\,(1-\mu)^{d-3\over 2}\,d\mu$ (it remains normalized, $\int_0^1{\cal
  D}\mu=1$). In the elastic case, the velocity distribution is Maxwellian.
Indeed, $\xi+\eta=1$ and $\lambda=0$ when $\epsilon=0$, and thence
$\Phi(x)=e^{-x/2}$.

Our primarily goal is to determine statistics of extremely fast
particles, namely the tail of the velocity distribution.  This can
be accomplished by noting that the large-$v$ behavior of the
velocity distribution is reflected by the small-$k$ behavior of
its Fourier transform. This is seen from the small-$x$ expansion
of the 1D solution (\ref{fsol}) which contains both regular and
singular terms: $\Phi(x)=1-{1\over 2}\,x+{1\over
3}\,x^{3/2}+\cdots$.  The dominant singular $x^{3/2}$ term
reflects the $w^{-4}$ tail of ${\cal P}(w)$. In general, an
algebraic tail of the velocity distribution (\ref{pscl}),
\begin{equation}
\label{tail}
{\cal P}(w)\sim w^{-\sigma} \qquad {\rm when} \qquad w\to\infty,
\end{equation}
indicates the existence of a singular component in the Fourier transform,
\begin{equation}
\label{sing}
\Phi_{\rm sing}(x)\sim x^{(\sigma-d)/2} \qquad {\rm when} \qquad x\to 0,
\end{equation}
and vice versa. To see this, it is convenient to take the Fourier
transform $\Phi(x)\propto \int_0^\infty dw\, w^{d-1}{\cal
P}(w)\,e^{iw\sqrt{x}}$, and write $x=-s^2$ thus recasting it into
the Laplace transform $I(s)\propto \int_0^\infty dw\,w^{d-1}{\cal
P}(w)\,e^{-ws}$. The small-$s$ expansion of the integral $I(s)$
contains regular and singular components.  For example, when
$\sigma<d$, the integral $I(s)$ diverges as $s\to 0$ and
integration over large-$w$ yields the dominant contribution
$I_{\rm sing}(s)\sim s^{\sigma-d}$.  When $d<\sigma<d+1$, $I(0)$
is finite, but the next term is the above singular term, so
$I(s)=I(0)+I_{\rm sing}(s)+\cdots$.  In general, the singular
contribution is $I_{\rm sing}(s)\sim s^{\sigma-d}$, thereby
leading to the singular term of Eq.~(\ref{sing}). This is of
course consistent with the one-dimensional behavior where
$\sigma=4$.

The exponent $\sigma$ can now be obtained by inserting
$\Phi(x)=\Phi_{\rm reg}(x)+\Phi_{\rm sing}(x)$ into Eq.~(\ref{fx}) and
equating the dominant singular terms. Combining $\Phi_{\rm reg}(0)=1$,
with the anticipated leading singular term of Eq.~(\ref{sing}), we
find that the exponent $\sigma$ is a root of the following integral
equation
\begin{equation}
\label{main}
1-\lambda\,{\sigma-d\over 2} =\int_0^1 {\cal
D}\mu\,\left[\xi^{(\sigma-d)/2}+\eta^{(\sigma-d)/2}\right].
\end{equation}
This equation has a trivial solution $\sigma=d+2$, following
from the identity $\int{\cal D}\mu\,(\xi+\eta)=1-\lambda$, where the
singular and the regular components simply coincide,
$x^{(\sigma-d)/2}=x$. Since we seek the leading {\em singular} term,
the solution of Eq.~(\ref{main}) must therefore satisfy $\sigma>d+2$.

The integral equation (\ref{main}) can be written explicitly in
terms of special functions
\begin{eqnarray}
\label{solve} &&1-\epsilon(1-\epsilon)\,{\sigma-d\over d}=\\
&&{}_2F_1 \left[{d-\sigma\over 2},{1\over2};{d\over 2};
1-\epsilon^2\right] +(1-\epsilon)^{\sigma-d}\,
{\Gamma\left({\sigma-d+1\over 2}\right)\Gamma\left({d\over
2}\right)\over \Gamma\left({\sigma\over
2}\right)\,\Gamma\left({1\over 2}\right)},\nonumber
\end{eqnarray}
with ${}_2F_1(a,b;c;z)$ the hypergeometric function \cite{aar}.
Interestingly, the exponent $\sigma\equiv\sigma(d,\epsilon)$ is a root
of the transcendental equation (\ref{solve}) and thence it depends in
a nontrivial fashion on spatial dimension $d$ and the restitution
coefficient $\epsilon$.

We first consider the dependence on the restitution coefficient by
considering the quasi-elastic limit $\epsilon\to 0$. As discussed
above, in the elastic case the velocity distribution is Maxwellian and
the Fourier transform is simply $\Phi(x)=e^{-x/2}$ \cite{caveat}. This
implies a diverging exponent $\sigma\to\infty$ as $\epsilon\to
0$. Therefore, the right-hand side of Eq.~(\ref{solve}) vanishes, and
the leading behavior is
\begin{equation}
\label{sig0} \sigma\simeq {d\over \epsilon}\qquad{\rm as}\qquad
\epsilon\to 0.
\end{equation}
One can further expand $\sigma(d,\epsilon)$ in the $\epsilon\to 0$
limit to find: $\sigma(d,\epsilon)=d\,
\epsilon^{-1}+a_1(d)\,\epsilon^{-1/2}+a_2(d)+\cdots$. We merely quote
the leading correction in the physically relevant spatial dimensions
$a_1(2)=-2(e^{-2}+1)/\sqrt{\pi}$ and $a_1(3)=-\sqrt{3\pi/2}$.
Clearly, the quasi-elastic limit is singular. Dissipation, even if
minute, seriously changes the nature of the system
\cite{bcdr,lcdkg,plmv}.

Next, we discuss the dependence on the dimension. First, one can
verify that $\sigma=4$ when $d=1$ using the identity
${}_2F_1(a,b;b;z)=(1-z)^{-a}$. The case $d=1$ is unique in that
the entire scaling function and in particular the exponent become
independent of the restitution coefficient. The case of
$d\to\infty$ is similar to the $\epsilon=0$ case in that the
inelastic nature of the collisions becomes irrelevant, the
velocity distribution is Maxwellian, and the exponent diverges,
$\sigma\to\infty$ as $d\to \infty$. In this limit, the second
integral in Eq.~(\ref{main}) is negligible as it vanishes
exponentially with the dimension. The first integral can be
evaluated by taking the limits $d\to\infty$ and $\mu\to 0$, with
$z=\mu d/2$ being finite. Then, the integration measure is
transformed ${\cal D}\mu\to (\pi z)^{-1/2}e^{-z}\,dz$, and
Eq.~(\ref{main}) becomes $1-\lambda u=\int_0^{\infty} dz (\pi
z)^{-1/2}e^{-[1+(1-\epsilon^2)u]z}$ where $u={\sigma\over d}-1$.
Performing the integration yields $1-\lambda
u=[1+(1-\epsilon^2)u]^{-1/2}$, from which we find $u$ and
\begin{equation}
\label{larged}
\sigma \simeq d\,{1+{3\over 2}\epsilon-\epsilon^3-\epsilon^{1/2}
\left(1+{5\over 4}\epsilon\right)^{1/2}\over \epsilon(1-\epsilon^2)},
\end{equation}
as $d\to\infty$.  In general, $\sigma\propto d$, and therefore,
the algebraic decay becomes sharper as the dimension increases.
The exponent $\sigma(d,\epsilon)$ increases monotonically with
increasing dimension, and additionally, it increases monotonically
with decreasing $\epsilon$ (see Fig.~1). Both features are
intuitive as they mirror the monotonic dependence of the energy
dissipation rate $\lambda=2\epsilon(1-\epsilon)/d$ on $d$ and
$\epsilon$.  Hence, the completely inelastic case provides a lower
bound for the exponent, $\sigma(d,\epsilon)\ge
\sigma(d,\epsilon=1/2)$ with $\sigma(d,1/2)=6.28753$, $8.32937$,
for $d=2$, $3$, respectively.  The former value should be compared
with $\sigma\approx 5$, obtained from numerical simulations
\cite{bmp}. The algebraic tails are characterized by unusually
large exponents which may be difficult to measure accurately in
practice (for typical granular particles $\epsilon\approx 0.1$
yielding $\sigma\approx 30$). Figure 1 also shows that the
quantity $\sigma/d$ weakly depends upon the dimension, and the
large-$d$ limit (\ref{larged}) provides a good approximation even
for small dimensions.

\begin{figure}
%\centerline{\epsfxsize=8cm \epsfbox{fig1.eps}} 
\centerline{\includegraphics[width=8cm]{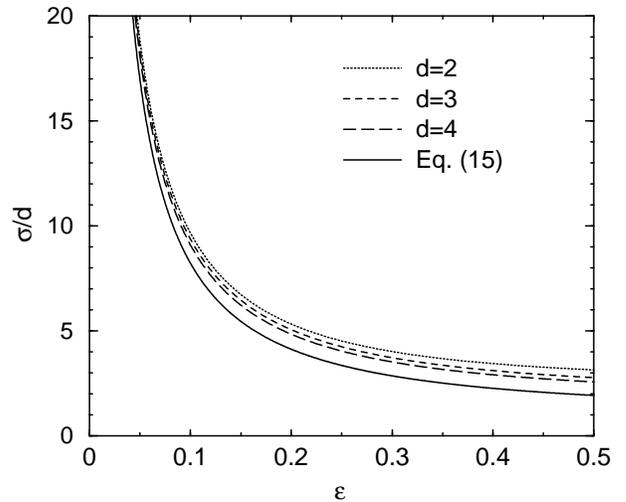}}
\caption{The exact exponent $\sigma$, obtained from Eq.~(\ref{main}), versus the
dissipation parameter $\epsilon$. The exponent was scaled by the
dimension $d$. Shown also is the limiting  large dimension
expression (\ref{larged}).}
\end{figure}

Thus far, we discussed only freely cooling systems where the energy decreases
indefinitely. In typical experimental situations, however, the system is
supplied with energy to balance the energy dissipation
\cite{lcdkg,ou,nbc,rm}.  Theoretically, it is natural to consider white noise
forcing \cite{ve}, i.e., coupling to a thermal heat bath which leads to a
nonequilibrium steady state. Specifically, we assume that in addition to
changes due to collisions, velocities may also change due to an external
forcing: \hbox{${dv_j\over dt}|_{\rm heat}=\xi_j$} with $j=1,\ldots,d$. We
use standard uncorrelated white noise \hbox{$\langle\xi_i(t) \xi_j(t')\rangle
  =2D\delta_{ij}\delta(t-t')$} with a zero average $\langle \xi_j\rangle=0$.
The temperature rate equation is modified by the additional source
term ${d\over dt}T+\lambda T^{3/2}=D$, and the system approaches a
steady state, $T_{\infty}=(D/\lambda)^{2/3}$. The relaxation
toward this state is exponential, $|T_{\infty}-T|\sim
e^{-t/\tau}$.

Uncorrelated white noise forcing amounts to diffusion in velocity
space, and Eq.~(\ref{fkt}) is modified as follows
${1\over \sqrt{T}}{\partial\over\partial t}\to {1\over \sqrt{T}}
{\partial\over\partial t}+Dk^2$. In
the steady state, the Fourier transform, $F({\bf
  k},t=\infty)\equiv \Psi(y)$ with $y=Dk^2$, obeys
\begin{equation}
\label{int}
(1+y)\,\Psi(y)=\langle \Psi(\xi y)\, \Psi(\eta y)\rangle,
\end{equation}
where integration with respect to the measure ${\cal D}\mu$ is
denoted by $\langle f\rangle =\int_0^1 {\cal D}\mu f(\mu)$.

Equation (\ref{int}) is solved recursively by employing the
cumulant expansion $\Psi(y)=\exp\left[\sum_{n\geq 1}
(-y)^nF_n\right]$.  Writing $1+y=\exp\left[\sum_{n\geq
1}(-y)^n/n\right]$, we recast Eq.~(\ref{int}) into
\begin{eqnarray}
\label{cumeqg}
\exp\left[-\sum_{n=1}^{\infty}{(-y)^n\over n}\right]=
\Bigg\langle\exp\left[-\sum_{n=1}^{\infty} (-y)^n G_n\right]\Bigg\rangle,
\end{eqnarray}
where $G_n=F_n(1-\xi^n-\eta^n)$.  The desired cumulants $F_n$ are
obtained by solving for $\langle G_n\rangle$ recursively and then
using $F_n=\langle G_n\rangle/\langle 1-\xi^n-\eta^n\rangle$.  In
one dimension $\langle \mu^n\rangle=1$ and one immediately obtains
$\langle G_n\rangle=n^{-1}$ \cite{bk}. In higher dimensions, the
averages acquire non-trivial dependence on $n$\cite{few} though
the qualitative nature of the solution remains the same since
$\langle G_n\rangle$ and hence the cumulants $F_n$ (as well as the
moments of the distribution) are positive.  This implies the
following small-$k$ behavior of the Fourier transform at the
steady state: $\ln F_{\infty}({\bf k})\sim -(k/k_0)^2$.
Consequently, the large velocity tail of the velocity distribution
is Gaussian
\begin{equation}
\label{std}
P_\infty(v)\sim e^{-(v/v_0)^2}\qquad {\rm when}\quad v\gg v_0.
\end{equation}
This behavior agrees with the prediction of Ref.~\cite{ccg} derived
via a small-$\epsilon$ expansion, but differs from the stretched
exponential behavior $\exp(-v^{3/2})$ found for the driven
inelastic hard sphere gas\cite{ve}.

We have studied inelastic gases within the framework of the
Boltzmann equation with a uniform collision kernel.  In the freely
evolving case, we have shown analytically that the density of
high-energy particles is suppressed algebraically.  The algebraic
tails are characterized by remarkably large exponents,
 and may be hard to distinguish from (stretched) exponential
 tails. Our results, combined with previous kinetic theory studies which
find exponential, stretched exponential, and Gaussian tails,
indicate that the extremal characteristics can be very sensitive
to parameters such as the restitution coefficient, and the
dimension\cite{ep,ve,jr}. These findings are all the more
intriguing when compared with molecular dynamics simulations and
scaling studies which suggest universal velocity distributions, at
least in some cases \cite{bcdr}. On the other hand, our results in
the forced case support the near-Maxwellian assumptions typically
used to obtain macroscopic transport coefficients from kinetic
theory \cite{sg}.

We thank A.~Baldassari for communicating the results of
Ref.~\cite{bmp} prior to publication, and for fruitful discussions. We
also thank G.~D.~Doolen and S.~Redner for their useful comments. This
research was supported by DOE (W-7405-ENG-36) and NSF(DMR9978902).

\end{multicols}

\begin{thebibliography}{99}
\bibitem{caveat}
     Every initial velocity distribution evolves towards the Maxwellian
     distribution if $d>1$ and $\epsilon=0$; for $d=1$, there is no
     relaxation in the elastic case.
\bibitem{few}
     The first three averages are $\langle G_1\rangle =1$, $\langle
     G_2\rangle={1\over 2}\langle G_1^2\rangle$, and $\langle
     G_3\rangle=\langle G_1G_2\rangle-{1\over 6}\langle G_1^3\rangle$.
\end{thebibliography}
\end{document}